\documentclass[12pt]{article}
\usepackage{graphicx}

\def\HEPHY{Institute for High Energy Physics, Austrian Academy of Sciences\\
Nikolsdorfergasse 18, A-1050 Vienna, Austria}
\def\SINP{D.V.~Skobeltsyn Institute of Nuclear Physics\\
 M.V.~Lomonosov Moscow State University, 119991, Moscow, Russia}
\def\RM3{Istituto Nazionale di Fisica Nucleare, Sezione Roma Tre\\
Via della Vasca Navale 84, I-00146 Roma, Italy}

\def\Title#1{\begin{center} {\Large #1 } \end{center}}
\def\Author#1{\begin{center}{ \sc #1} \end{center}}
\def\Address#1{\begin{center}{ \it #1} \end{center}}

\newenvironment{Abstract}{\begin{quotation}  }{\end{quotation}}
\newenvironment{Presented}{\begin{quotation} \begin{center} 
             PRESENTED AT\end{center}\bigskip 
      \begin{center}\begin{large}}{\end{large}\end{center} \end{quotation}}

\def\beq{\begin{equation}}
\def\eeq#1{\label{#1}\end{equation}}
\def\eeqn{\end{equation}}

\def\beqa{\begin{eqnarray}}
\def\eeqa#1{\label{#1}\end{eqnarray}}
\def\eeqan{\end{eqnarray}}

\let\bar=\overbar

\def\Dslash{\not{\hbox{\kern-4pt $D$}}}
\def\dslash{\not{\hbox{\kern-2pt $\del$}}}

\def\msb{{\bar{\ssstyle M \kern -1pt S}}}

\begin{document}
\begin{titlepage}

\vfill
\Title{QCD sum-rule results for heavy-light meson decay constants\\[2mm]
         and comparison with lattice QCD}
\vfill
\Author{Wolfgang Lucha}
\Address{\HEPHY}
\Author{Dmitri Melikhov}
\Address{\HEPHY}
\Address{\SINP}
\Author{Silvano Simula}
\Address{\RM3}
\vfill
\begin{Abstract}
Updated predictions for the decay constants of the $D$, $D_s$, $B$ and $B_s$ mesons obtained from Borel QCD sum rules for heavy-light currents are presented and compared with the recent lattice averages performed by the Flavor Lattice Averaging Group. An excellent agreement is obtained in the charm sector, while some tension is observed in the bottom sector. Moreover, available lattice and QCD sum-rule calculations of the decay constants of the vector $D^*$, $D_s^*$, $B^*$ and $B_s^*$ mesons are compared. Again some tension in the bottom sector is observed.
\end{Abstract}
\vfill
\begin{Presented}
the 8$^{th}$ International Workshop on\\ the CKM Unitarity Triangle (CKM 2014)\\
Vienna, Austria,  September 8--12, 2014
\end{Presented}
\vfill
\end{titlepage}
\def\thefootnote{\fnsymbol{footnote}}
\setcounter{footnote}{0}


\section{Leptonic decay constants from QCD sum rules}

Leptonic decay constants of pseudoscalar (PS) heavy-light mesons are crucial hadronic ingredients relevant for the extraction of the Cabibbo-Kobayashi-Maskawa (CKM) matrix elements from the experimental data on the weak decays of a heavy-light meson $H$ to a lepton-neutrino pair via flavor-changing transitions \cite{Rosner:2013ica}, $H \to \ell \nu_\ell$, and on the rare leptonic decays of neutral PS mesons to a charged-lepton pair via flavor-changing neutral currents \cite{Aaij:2012nna}, $H \to \ell^+ \ell^-$.
Moreover, the leptonic decay constants of vector (V) heavy-light mesons are relevant quantities in the heavy-quark phenomenology, like, e.g., for describing the contributions of vector poles coupled to weak currents mediating the semileptonic decays of PS heavy-light mesons. 

Within the method of QCD sum rules (QCD-SR) \cite{Shifman:1978bx,Aliev:1983ra} the extraction of the leptonic decay constants of ground-state PS and V mesons is based on the analysis of the two-point correlation functions
 \begin{eqnarray}
     \label{eq:PS} 
     i \int d^4x e^{ip \cdot x} \langle 0| T\left[j_5(x) j_5^\dagger(0) \right] |0 \rangle & = & \Pi^{PS}(p^2) ~ , \\
     \label{eq:V}
     i \int d^4x e^{ip \cdot x} \langle 0| T\left[j_\nu(x) j_{\nu^\prime}^\dagger(0) \right] |0 \rangle & = & \left( -g_{\nu \nu^\prime} + \frac{p_\nu p_{\nu^\prime}}{p^2} \right)
               \Pi^V(p^2) + \frac{p_\nu p_{\nu^\prime}}{p^2} \Pi_L^V(p^2) ~ ,
 \end{eqnarray}
where $j_5(x) = (m_h + m_q) \overline{q}(x) i \gamma_5 h(x)$ and $j_\nu(x) = \overline{q}(x) \gamma_\nu h(x)$ are interpolating heavy-light quark currents with $h = c, b$ and $q = u, d, s$.

The correlators $\Pi^{PS}(p^2)$ and $\Pi^V(p^2)$ have both a hadronic and a quark-gluon (OPE) representation.
After Borelization one has:
  \begin{eqnarray}
     \label{eq:OPE}
     \Pi^{PS(V)}(\tau) & = & f_{PS(V)}^2 M_{PS(V)}^{4(2)} ~ e^{-M_{PS(V)}^2 \tau} + \int_{s_{phys}^{PS(V)}}^\infty ds ~ e^{-s \tau} \rho_{hadron}^{PS(V)}(s) 
                                          \nonumber \\
                                & = & \int_{(m_h + m_q)^2}^\infty ds ~ e^{-s \tau} \rho_{pert}^{PS(V)}(s, \mu) + \Pi_{power}^{PS(V)}(\tau, \mu) ~ ,                          
 \end{eqnarray} 
where $f_{PS(V)}$ is the leptonic decay constant\footnote{The leptonic decay constants are defined as: $f_{PS} M_{PS}^2 \equiv \langle 0| j_5(0) | H \rangle$ and $f_V M_V \varepsilon_\nu^V \equiv \langle 0| j_\nu(0) | H^* \rangle$, where $\varepsilon_\nu^V$ is the polarization vector of the $H^*$ meson.}, $M_{PS(V)}$ the mass of the ground state, $s_{phys}^{PS(V)}$ the threshold for excited states, $\rho_{pert}^{PS(V)}$ the perturbative spectral density, $\Pi_{power}^{PS(V)}$ the power corrections containing the contributions of all vacuum condensates and $\mu$ is the subtraction point introduced by the OPE.

The perturbative spectral density $\rho_{pert}^{PS(V)}$ can be expanded as a series of powers of the strong coupling constant $\alpha_s(\mu)$ and it has been calculated beyond the leading order (LO), given by the simple heavy-light loop, by including two- and three-loop contributions.
The NLO corrections, originating from two loops related to gluon exchanges, are known from Refs.~\cite{Broadhurst:1991fc, Generalis:1990id} in the case of the PS channel and from Ref.~\cite{Dominguez:1990nz} in the case of the V one. 
The NNLO contributions for both PS and V correlators have been calculated in Ref.~\cite{Chetyrkin:2001je} in the case of massless light quarks.
 The (small) corrections for finite values of the light-quark mass are known from Ref.~\cite{Jamin:2001fw} for the PS correlator and only very recently from Ref.~\cite{Gelhausen:2013wia} for the V one.
An important result is that the perturbative series expressed in terms of the heavy-quark pole mass is not convergent at all, as shown in Refs.~\cite{Jamin:2001fw, Lucha:2011zp, Lucha:2013gta} for the PS correlator and in Ref.~\cite{Lucha:2014xla} for the V one.
This problem can be cured by rearranging the perturbative expansion in terms of the running $\overline{\rm MS}$ mass, which, at variance with the pole mass, is also free from renormalon ambiguities.

As for the power corrections $\Pi_{power}^{PS(V)}$, the contributions from vacuum condensates up to dimension $d = 6$, namely the quark ($d = 3$), gluon ($d = 4$), quark-gluon ($d = 5$) and four-quark condensates ($d = 6$), are known at LO.
The only NLO correction available has been calculated for the quark-condensate contribution in Ref.~\cite{Jamin:2001fw} for the PS channel and in Ref.~\cite{Gelhausen:2013wia} for the V one.

The explicit expressions for $\rho_{pert}^{PS(V)}$ and $\Pi_{power}^{PS(V)}$ in terms of the running $\overline{\rm MS}$ mass can be found in Refs.~\cite{Gelhausen:2013wia, Lucha:2014xla}.

 The Borelized correlator $\Pi^{PS(V)}(\tau)$ is dominated by the ground-state contribution at large values of the Borel variable $\tau$, where, however, any truncated OPE is not expected to converge. 
 An effective tool for eliminating the contribution of excited states at intermediate values of $\tau$ is the quark-hadron duality introduced for the first time in Ref.~\cite{Shifman:1978bx}, namely
  \begin{equation}
      \int_{s_{phys}^{PS(V)}}^\infty ds ~ e^{-s \tau} \rho_{hadron}^{PS(V)}(s) = \int_{s_{eff}^{PS(V)}}^\infty ds ~ e^{-s \tau} \rho_{pert}^{PS(V)}(s, \mu) ~ ,
      \label{eq:duality}
  \end{equation}
where $s_{eff}^{PS(V)}$ is an effective threshold.
Therefore, the leptonic decay constant of the ground state can be easily extracted from the {\it dual} correlator $\Pi_{dual}^{PS(V)}(\tau)$ given by
 \begin{eqnarray}
    \Pi_{dual}^{PS(V)}(\tau) & = & \int_{(m_h + m_q)^2}^{s_{eff}^{PS(V)}(\tau, \mu)} ds ~ e^{-s \tau} \rho_{pert}^{PS(V)}(s, \mu) + 
                                                     \Pi_{power}^{PS(V)}(\tau, \mu) \nonumber \\ 
                                           & = & f_{PS(V)}^2 M_{PS(V)}^{4(2)} ~ e^{-M_{PS(V)}^2 \tau} ~ .
    \label{eq:dual}
 \end{eqnarray}

The extraction procedure starts with the choice of the Borel window $\tau_{min} \leq \tau \leq \tau_{max}$, where the constraint $\tau \geq \tau_{min}$ guarantees that the ground state provides a sizable contribution (typically $> 50 \%$) to the full correlator $\Pi^{PS(V)}(\tau)$, while the constraint $\tau \leq \tau_{max}$ keeps the power corrections sufficiently small numerically.

Then, the effective threshold $s_{eff}^{PS(V)}$ is chosen by requiring that in the given Borel window the dual mass $M_{dual}^{PS(V)}(\tau)$, defined as the logarithmic slope of the dual correlator $\Pi_{dual}^{PS(V)}(\tau)$, reproduces the experimental meson mass $M_H$.

Any deviation of $M_{dual}^{PS(V)}(\tau)$ from $M_H$ means that contaminations from excited states are present in the chosen Borel window.
A strategy, commonly adopted in literature, is to assume that the effective threshold $s_{eff}^{PS(V)}$ is a constant, independent of $\tau$.
However, the crucial point is that the effective threshold $s_{eff}^{PS(V)}$, appearing in the duality relation (\ref{eq:duality}), is in general a function of the Borel variable $\tau$.
Such a feature has been pointed out for the first time in Ref.~\cite{LMS_1}, where the case of exactly solvable potential models was investigated in detail, and then applied to QCD in Refs.~\cite{Lucha:2011zp, Lucha:2013gta, Lucha:2014xla, LMS_2}.
The $\tau$-dependence of $s_{eff}^{PS(V)}$ reduces the deviations of $M_{dual}^{PS(V)}(\tau)$ from $M_H$, which implies that the contaminations from the excited states are reduced in the Borel window, leading to a clear, important improvement of the quality of the dual correlator $\Pi_{dual}^{PS(V)}(\tau)$.

In what follows we will limit ourselves only to the Borel QCD-SR results for the leptonic decay constants obtained adopting a $\tau$-dependent effective threshold, {\it i.e.}~to the results of Refs.~\cite{Lucha:2011zp, Lucha:2013gta, Lucha:2014xla}, where a polynomial Ansatz is used to parameterize $s_{eff}^{PS(V)}(\tau)$, and to those of Ref.~\cite{Gelhausen:2013wia}, where the quantities $s_{eff}^{PS(V)}(\tau_i)$ for each point $\tau_i$ of the (discretized) Borel window are treated as free parameters\footnote{Note that the calculation of $M_{dual}^{PS(V)}$ requires the knowledge of the derivative of $s_{eff}^{PS(V)}(\tau)$. This effect is ignored in Ref.~\cite{Gelhausen:2013wia}.}.


\section{Charm sector}

For the OPE input parameters, namely the light and heavy quark masses, the condensates and the strong coupling $\alpha_s(M_Z)$ required at the QCD level, a rather standard set of values is commonly adopted in the literature (see, {\it e.g.}, Table 1 of Ref.~\cite{Gelhausen:2013wia}). 

\begin{figure}[htb]
\centering
\includegraphics[height=5.0cm]{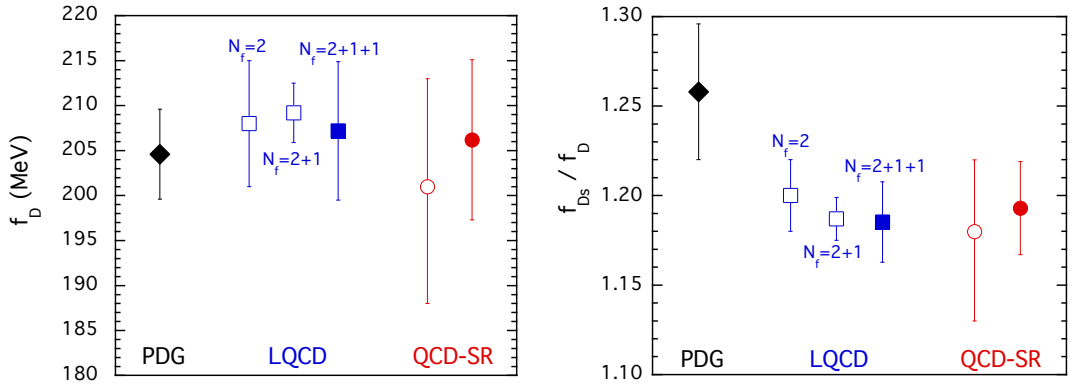}
\caption{\small \it QCD-SR results for $f_D$ (left) and $f_{D_s} / f_D$ (right) obtained in Ref.~\cite{Gelhausen:2013wia} (open dots) and in Ref.~\cite{Lucha:2011zp} (full dots), compared with the corresponding PDG values \cite{PDG} (diamonds) and with the recent LQCD averages from FLAG \cite{FLAG} (open and full squares) carried out at various values of the number of dynamical sea quarks ($N_f = 2, 2+1, 2+1+1$) considered in the lattice QCD simulations.}
\label{fig:fDfDs}
\end{figure}

The QCD-SR results for $f_D$ and $f_{D_s} / f_D$ obtained in Refs.~\cite{Gelhausen:2013wia} and \cite{Lucha:2011zp} are shown in Fig.~\ref{fig:fDfDs} and compared with the corresponding PDG values \cite{PDG} and with the averages of lattice QCD (LQCD) simulations with $N_f = 2, 2+1, 2+1+1$ dynamical quarks, provided recently by the Flavor Lattice Averaging Group (FLAG) \cite{FLAG}.
It can be seen that an excellent agreement exists between LQCD and QCD-SR calculations (where in the latter the PDG value of the charm mass, $\overline{m}_c(\overline{m}_c) = 1.275 (25)$ GeV, has been considered) as well as with the PDG value for $f_D$.
Only a moderate tension (at $1.5 \div 2 ~ \sigma$ level) is visible in the case of the PDG ratio $f_{D_s} / f_D$.

\begin{figure}[htb]
\centering
\includegraphics[height=5.0cm]{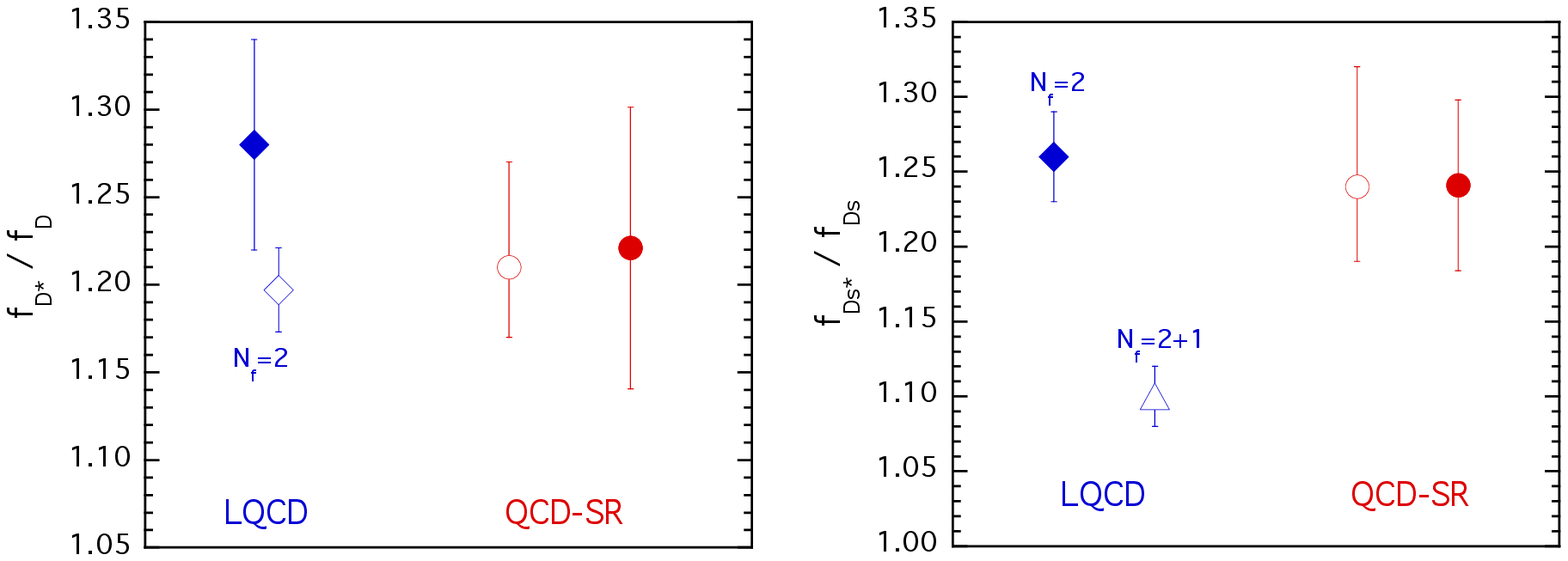}
\caption{\small \it QCD-SR results for $f_{D^*} / f_D$ (left) and $f_{D_s^*} / f_{D_s}$ (right) obtained in Ref.~\cite{Gelhausen:2013wia} (open dots) and in Ref.~\cite{Lucha:2014xla} (full dots), compared with the LQCD results of Ref.~\cite{Becirevic:2012ti} (full diamonds), Ref.~\cite{Becirevic:2014kaa} (open diamond) and Ref.~\cite{Donald:2013sra} (triangle).}
\label{fig:fDstar}
\end{figure}

The QCD-SR results for the ratios $f_{D^*} / f_D$ and $f_{D_s^*} / f_{D_s}$ obtained in Refs.~\cite{Gelhausen:2013wia} and \cite{Lucha:2014xla} are shown in Fig.~\ref{fig:fDstar} and compared with the LQCD results with $N_f = 2$ and $N_f = 2+1$ from Refs.~\cite{Becirevic:2012ti, Becirevic:2014kaa, Donald:2013sra}.
Again a good agreement is observed with the exception of the LQCD result of Ref.~\cite{Donald:2013sra} with $N_f = 2+1$.

\section{Beauty sector}

As pointed out in Ref.~\cite{Lucha:2013gta}, the decay constant $f_B$ extracted from QCD-SR is very sensitive to the input value of the b-quark mass, $\overline{m}_b(\overline{m}_b)$. 
The origin of the above sensitivity is not surprising and it can be understood easily even in non-relativistic quantum-mechanical models.
In QCD one finds that $\delta f_B \simeq -0.37 ~ \delta \overline{m}_b(\overline{m}_b)$ \cite{Lucha:2013gta}.
The above feature opens the possibility to determine $\overline{m}_b(\overline{m}_b)$ using as input a precise value for $f_B$, like the one that can be provided by LQCD.

\begin{figure}[htb]
\centering
\includegraphics[height=5.0cm]{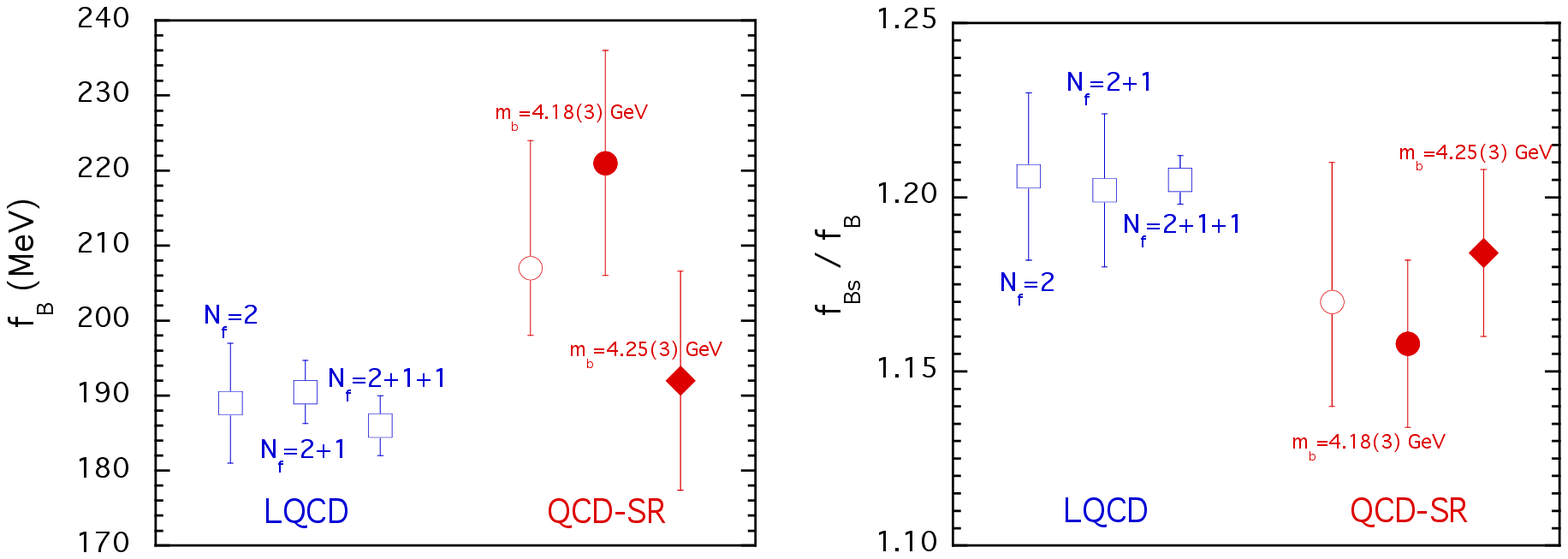}
\caption{\small \it QCD-SR results for $f_B$ (left) and $f_{B_s} / f_B$ (right) obtained in Ref.~\cite{Gelhausen:2013wia} (open dots) and in Ref.~\cite{Lucha:2013gta} (full dots and diamonds), compared with the FLAG averages \cite{FLAG} (open squares) of LQCD calculations with $N_f = 2, 2+1, 2+1+1$ dynamical quarks. The open and full dots have been obtained using the PDG value for the $b$-quark mass, $\overline{m}_b(\overline{m}_b) = 4.18 (3)$ GeV \cite{PDG}, while the diamonds correspond to $\overline{m}_b(\overline{m}_b) = 4.25 (3)$ GeV, obtained in \cite{Lucha:2013gta} using the FLAG averages as input for $f_B$. }
\label{fig:fBfBs}
\end{figure}

As shown in Fig.~\ref{fig:fBfBs}, the FLAG averages for $f_B$ indicate a central value around $190$ MeV with an error of $\simeq 2 \%$.
Using the PDG value for the b-quark mass, $\overline{m}_b(\overline{m}_b) = 4.18 (3)$ GeV \cite{PDG}, the QCD-SR results for $f_B$ from Refs.~\cite{Gelhausen:2013wia} and \cite{Lucha:2013gta} are $\simeq 10 \div 15 \%$ higher than the FLAG averages, with a total uncertainty of $\simeq 6 \%$.
On the other hand, when the FLAG average is used as input for $f_B$, the $b$-quark mass obtained from QCD-SR is $\overline{m}_b(\overline{m}_b) = 4.25 (3)$ GeV \cite{Lucha:2013gta}, {\it i.e.}~$\simeq 1.5 \%$ higher than the PDG value, with an uncertainty of $\simeq 0.7 \%$. 
Therefore, the above findings suggest that a moderate tension (at the $1.5 \div 2\sigma$ level) occurs between the LQCD determinations of $f_B$ and the QCD-SR results for $\overline{m}_b(\overline{m}_b)$.

\begin{figure}[htb]
\centering
\includegraphics[height=5.0cm]{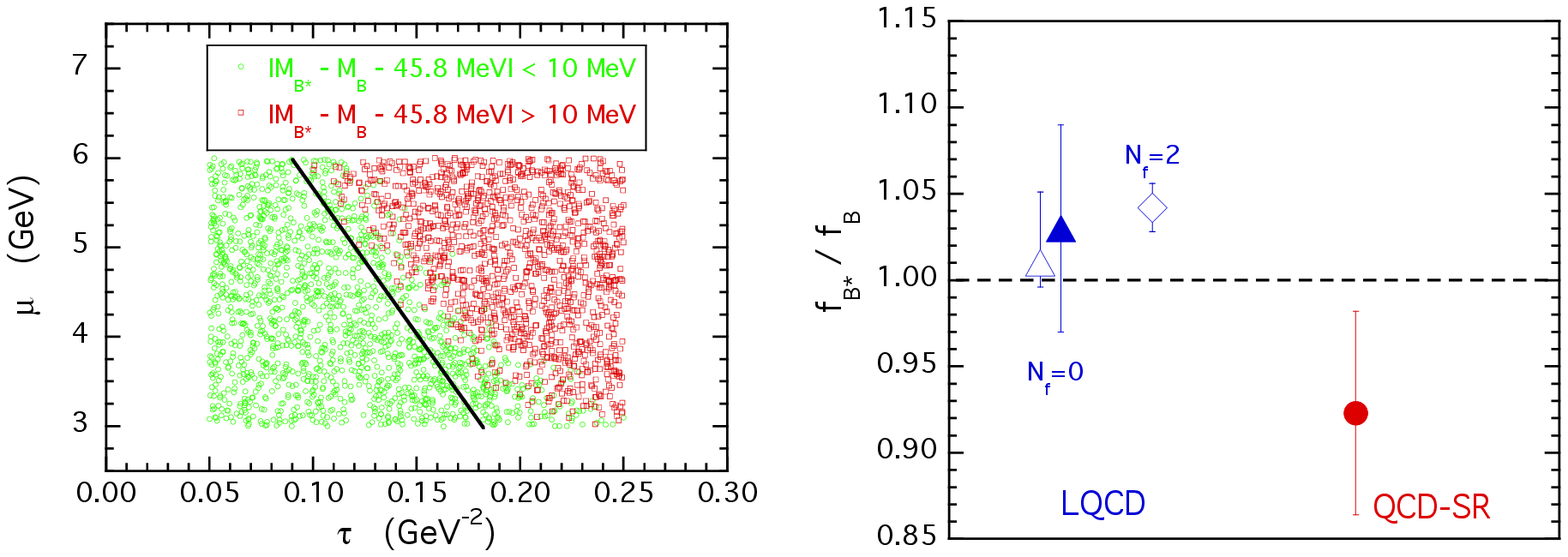}
\caption{\small \it Left panel: the $(\mu, \tau)$ plane of the OPE parameter space, where green (red) markers correspond to the cases in which the {\it dual} $B^*$-$B$ mass splitting deviates from its experimental value, equal to $45.8$ MeV \cite{PDG}, less (more) than $10$ MeV. Right panel: QCD-SR result for the ratio $f_{B^*} / f_B$ from Ref.~\cite{Lucha:2014nba} (dot) compared with the LQCD results obtained in Ref.~\cite{Bernard:2001fz} (open triangle), Ref.~\cite{Bowler:2000xw} (full triangle) and Ref.~\cite{Becirevic:2014kaa} (diamond).}
\label{fig:fBstar}
\end{figure}

In the case of the QCD-SR for the $B^*$ meson a new problem emerges.
Even if a $\tau$-dependent effective threshold is adopted, the dual mass splitting $(M_{dual}^{B^*} - M_{dual}^B)$ does not reproduce the experimental mass splitting $(M^{B^*} - M^B) = 45.78 (35)$ MeV \cite{PDG} in the full OPE parameter space, as is clearly illustrated in the left panel of Fig.~\ref{fig:fBstar}. 
This problem can be cured only by assuming that the Borel window, in particular the upper limit $\tau_{max}$, depends on the subtraction point $\mu$ (see the solid line in the left panel of Fig.~\ref{fig:fBstar}).

In this way a strong sensitivity of $f_{B^*}$ to the subtraction point is found \cite{Lucha:2014nba}.
In particular, the ratio $f_{B^*} / f_B$ turns out to be definitely below unity for $\mu > 3$ GeV and above unity only for $\mu < 3$ GeV, where, however, the hierarchy of the various perturbative orders is lost even using the running $\overline{\rm MS}$ mass.
Therefore, taking into account a range of values of $\mu$ from $3$ to $6$ GeV, we obtain \cite{Lucha:2014nba}  
 \begin{equation}
     f_{B^*} / f_B = 0.923 ~ (59) ~ , \qquad \qquad f_{B_s^*} / f_{B_s} = 0.932 (47) ~ .
     \label{eq:fBstar}
 \end{equation}

Our finding (\ref{eq:fBstar}) for the ratio $f_{B^*} / f_B$ is reported in the right panel of Fig.~\ref{fig:fBstar} and compared with few available LQCD results, which, however, suggest a value of the ratio above unity.


\section{Conclusions}

In this contribution we have highlighted the important improvements in the quality of Borel QCD-SR predictions for the leptonic decay constants of heavy-light mesons achieved during the last years thanks to a better knowledge of the perturbative spectral density and of the condensate contributions as well as to new algorithms that allow a significative reduction of the excited-state contaminations in the dual correlators for both PS and V heavy-light currents.

In the charm sector there is an excellent agreement between the predictions of LQCD, as analyzed by FLAG, and those from QCD-SR, adopting in the latter the PDG value for the charm quark mass.

In the beauty sector a moderate tension (at the $1.5 \div 2\sigma$ level) occurs between the FLAG averages for $f_B$ and the QCD-SR result for the $b$-quark mass.
Moreover, the reproduction of the $B^*$-meson mass is problematic in some parts of the OPE parameter space and a $\mu$-dependent Borel window has to be considered to guarantee the reproduction of the experimental $B^*$-$B$ mass splitting.
While available LQCD predictions for the ratio $f_{B^*} / f_B$ suggest a value above unity, the Borel QCD-SR is remarkably sensitive to the value of $\mu$ and favors values of the ratio $f_{B^*} / f_B$ below unity in the range $3 < \mu ~ (\rm{GeV}) < 6$.

The presence of the above tensions in the beauty sector and their absence in the charm one are open issues to be further investigated.



\end{document}